\documentstyle [12pt]{article}
\begin{document}
\setcounter{page}1
\setcounter{footnote}0

\begin{flushright}
\large\bf
KIPT E96-2
\end{flushright}

\begin{center}
{\large \bf
National Science Center\\
"Kharkov Institute of Physics and Technology"\\
\vspace{5cm}
M.I.Ayzatsky\footnote{ M.I.Ayzatsky (N.I.Aizatsky)\\
National Science Center
"Kharkov Institute of Physics and Technology"\\
Akademicheskaya 1,
Kharkov, 310108, Ukraine\\
e-mail:aizatsky@nik.kharkov.ua}\\
\vspace{2cm}
ANALYTICAL SOLUTIONS IN THE TWO-CAVITY COUPLING PROBLEM  \\
\vspace{1.5cm}
E-Preprint\\
\vspace{7cm}
Kharkov --- 1996}
\end{center}
\newpage

\begin{abstract}
Analytical solutions of precise equations that describe the rf-coupling
of two cavities through a co-axial cylindrical hole are given for
various limited cases. For their derivation we have used the method of
solution of an infinite set of linear algebraic equations, based on its
transformation into dual integral equations.
\end{abstract}

\section{Introduction}

In the papers \cite{r1,r2,r3}, we derived precise equations, describing
the rf-coupling of two cavities through a centerhole of arbitrary dimensions.
On the base of these equations we numerically calculated the relationship
of coupling coefficients versus different parameters
(frequency, hole radius, etc.).
This paper presents analytical solutions of these equations for various
limited cases. In particular, it is explicitly shown that in the case of small
holes $(a\rightarrow 0)$ the formulated equations agree with those
derived in the papers \cite{r4,r5,r6} on the base of quasi-static approach.
Besides, expressions are derived for coupling coefficients which are valid up
to the second order in the relation of the hole dimension ($a$) with the
free-space wavelength ($\lambda$). For derivation of these expressions we
have used the method of solving of an infinite set of linear algebraic
equations, based on its transformation into dual integral equations.

\section{Problem definition. Original equations}

Let us consider the coupling of two cavities through a circular hole
with the radius~$ \; a \; $ in a separating wall that has
the thickness~$ \; t$. For simplicity's sake, we will consider the case of
two identical cavities, with  $ \ b\ $-being the cavity radii and $\ d $ ---
their length. In the papers  \cite{r1,r2,r3} it was demonstrated that
if the field is expanded with the short-circuit resonant cavity modes
and  $E_{010}$-modes are selected as fundamental, the precise set of
equations will consist of two equations for the amplitudes of $E_{010}$-modes,
where coupling coefficients are defined by the way of solution of an
infinite set of linear algebraic equations. Let us generalize the case
considered in  \cite{r1,r2,r3}, choosing as fundamental $E_{0qp}$-modes
of closed cavities ($q$ is the number of field variations across the radius,
$p$ is the number of field variations along the longitudinal coordinates).
Using the method, similar to the one in \cite{r1,r2,r3}, one can show that
the set of equations, describing the system under consideration, has the form;
\begin{eqnarray}
\epsilon_pZ_{q,p}a_{q,p}^{(1)}=-\omega_{q,p}^{2}\frac{4}{3\pi
J_1^2(\lambda_q)} \frac{a^3}{b^2d}
\left[a_{q,p}^{(1)}\Lambda_1-(-1)^p a_{q,p}^{(2)}\Lambda_2\right]\label{eq1}\\
 \epsilon_pZ_{q,p}a_{q,p}^{(2)}=-\omega_{q,p}^{2}\frac{4}{3\pi
J_1^2(\lambda_q)} \frac{a^3}{b^2d}
\left[a_{q,p}^{(2)}\Lambda_1-(-1)^p a_{q,p}^{(1)}\Lambda_2\right] \label{eq2}
\end{eqnarray}
where
$$
Z_{q,p}=\omega_{q,p}^2-\omega^2,\
\omega_{q,p}^2=c^2\left[\lambda_q^2/b^2+ \left(p\pi/d\right)^2\right],
$$
$$
\epsilon_p=
\left\{
\begin{array}{c}
2,p=0 \\
1,p\neq 0
\end {array} ,
\right. \; p=0,1,...\infty,
\; J_0(\lambda_s)=0,\; s=1,2,...\infty ,
$$
$a_{q,p}^{(i)}$ is the amplitude of $E_{0qp}$-mode in the $i$-th cavity
$(i=1,2)$.
The normalized coupling coefficients  $\Lambda_i$ are determined
by the expression:
\begin{equation}
\Lambda_i=\Lambda_i(\omega)=J_0^2\left(\theta_q
\right)\sum_{s=1}^\infty w_s^{(i)}/ \left(\lambda_s^2-\theta_q^2
\right), \label{eq3}
\end{equation}
where $w_s^{(i)}$ are the solution of the following set of linear equations:
\begin{eqnarray}
w_m^{(1)}+\sum_sG_{m,s}\left(w_s^{(1)}f_m^{(1)}+w_s^{(2)}f_m^{(2)}\right)
=3\pi f_m^{(1)}/\left(\lambda_m^2-\Omega_\ast^2\right),
\label{eq4} \\
w_m^{(2)}+\sum_sG_{m,s}\left(w_s^{(2)}f_m^{(1)}+w_s^{(1)}f_m^{(2)}\right)
=3\pi f_m^{(2)}/\left(\lambda_m^2-\Omega_\ast^2\right),
\label{eq5}
\end{eqnarray}
$$
f_m^{(j)}= \frac{\mu_m}{sh(q_m)} \left\{
\begin{array}{lr}
ch(q_m)-ch(q_m t/l),& j=1 \\
ch(2q_m d/l)-1, & j=2,
\end{array}
\right.
$$
$$
q_m=\mu_ml/a, \ l=2d+t, \ \mu_m=\sqrt{\lambda_m^2-\Omega^2}, \ \Omega=
\omega a/c,\ \Omega_\ast^2=\Omega^2-\pi^2 a^2 p^2/d^2.
$$
\begin{eqnarray}
G_{m,s}=B_{m,s}-\frac {1}{2\mu_m} \delta_{m,s} cth(\frac {d}{a} \mu_m)+
\nonumber \\
+\frac{2 \pi a^2 \theta_q^3 J_0^2(\theta_q)}
{d b \ \epsilon_p \chi_q (\lambda_m^2-\theta_{q}^2)
(\lambda_s^2-\theta_{q}^2)(\mu_m^2+\pi^2a^2p^2/d^2)}, \label{eq6}
\end{eqnarray}
\begin{equation}
B_{m,s}=\pi \frac {a} {b} \sum_{\ell=1}^{\infty}
\frac{ \theta_{\ell}^2 J_0^2(\theta_{\ell}) R_{\ell}}
{\chi_{\ell}(\lambda_m^2-\theta_{\ell}^2) (\lambda_s^2-\theta_{\ell}^2)},
\label{eq7}
\end{equation}
$$
\theta_{\ell}=\lambda_{\ell}a/b, \;
\chi_{\ell}=\pi\lambda_{\ell} J_1^2(\lambda_{\ell})/2, \;
\nu_{\ell}=\sqrt{\theta_{\ell}^2-\Omega^2},
$$
\begin{equation}
R_{\ell}=\left\{
\begin{array}{lr}
\theta_q cth(\nu_q d/a)/\nu_q -2a\theta_q/\left\{\epsilon_p d
\left(\nu_q^2+\pi^2a^2p^2/d^2\right)\right\},& \ \ell=q, \\
\theta_{\ell}cth(\nu_{\ell} d/a)/\nu_{\ell},& \ \ell \neq q.
\end{array} \right. \label{eq8}
\end{equation}

The coefficients $w_s^{(i)}$ have a simple physical sense. Really, it is easy
to show that the tangential electric field component in the left
cross-section of the coupling hole $E_r^{(-)}(r)$ has the form
\begin{equation}
E_r^{(-)}(r)=
E_{ind}^{(1)}(r)-E_{ind}^{(2)}(r)=\tilde E_{0,q,p}^{(1)} Q^{(1)}(r)-
\tilde E_{0,q,p}^{(2)}Q^{(2)}(r), \label{eq9}
\end{equation}
where
$$
Q^{(i)}(r)=\frac{1}{3\pi}\sum_s
\frac{J_1\left(\lambda_s r/a\right)}{J_1\left(\lambda_s \right)}
w_s^{(i)},
$$
$\tilde E_{0,p,q}^{(1)}$ is the value of the longitudinal (perpendicular
to the hole) electric field of
$(0,q,p)$-mode in the first cavity on the left coupling hole cross-section
at  $r=a$, while  $\tilde E_{0,p,q}^{(2)}$ is the same value for the
right-hand cavity on the right coupling hole cross-section
at  the same radius.

From the expression (\ref{eq9}) it follows that the tangential electric field
component on the left coupling hole cross-section\footnote{The same is true
for the right cross section} is equal to the difference of two induced fields,
each of which is proportional to the perpendicular electric field components
of  $E_{0,q,p}$-modes, taken to be fundamental. There, the coefficients
$w_s^{(i)}$  are the ones of expansion of the appropriate functions
with the complete set of functions
$\left\{J_1 \left(\lambda_s r/a\right)\right\}$.

Note that the coefficients  $w_m^{(i)}$  can be re-defined which will cause
changes in Eqs.(\ref{eq4},\ref{eq5}), in the form of relationship
(\ref{eq9}), and, consequently, in the above nature of proportionality.
For example, in (\ref{eq9}) one can obtain proportionality  $E_{ind}^{(i)}$
to the longitudinal electric field component of $(0,q,p)$-mode at $r=0$.
While defining $w_m^{(i)}$, we proceeded from the following condition:
for the two different cavities, in the case $t=0$ (infinitely thin wall)
and taking $E_{0,q,0}$-modes to be fundamental:
$Q^{(1)}(r)=Q^{(2)}(r)=Q(r)$, then we have in this case
$E_r^{(-)}(r)=E_r^{(+)}(r)=Q(r)\left( \tilde E_{0,q,0}^{(1)}-\tilde
E_{0,q,0}^{(2)}\right)$. This condition determines the tangential electric
field component on the hole while having  $E_{0,q,p}$-modes as fundamental.
With such a normalization of $w_m^{(i)}$, their determining set
of infinite linear equations (\ref{eq4},\ref{eq5}) acquires the most
symmetrical form.

Thus, the two-cavity coupling problem, rigorously formulated on the base
of the electric field expansion with the short-circuit resonant cavity mode,
is reduced to the induced field definition on the right and left
cylindrical hole cross-section.

\section{Infinitely thin wall case}

An important role in the problem of cavity coupling plays the case
of infinitely thin wall, dividing the cavities $(t=0)$.
In this case, from  Eqs.(\ref{eq4},\ref{eq5}) it follows that
$w_m^{(1)}=w_m^{(2)}=w_m$.  In this case the set of equations for  $w_m$
will take on the form\footnote{We have neglected terms of order
$a^5$ in the expression (\ref{eq6}) for $G_{m,s}$}:
\begin{equation}
\sum_{s=1}^\infty w_s B_{m,s}
=3\pi/\left\{2\left(\lambda_m^2-\Omega_\ast^2\right)\right\}.
\label{eq10}
\end{equation}
For the case $t=0$ \ $\Lambda_1=\Lambda_2=\Lambda$, where
\begin{equation}
\Lambda=J_0^2\left(\theta_q\right)\sum_s
w_s/\left(\lambda_s^2-\theta_q^2\right). \label{eq11}
\end{equation}

\subsection{Small coupling hole case $(a\rightarrow 0)$}

If in Eqs.(\ref{eq10},\ref{eq11}) the hole radius
tends to zero\footnote{In this case, as follows from
Eqs.(\ref{eq1},\ref{eq2}), the coupling coefficients will be proportional
to $a^3$}, then Eq.(\ref{eq10}) will become:
\begin{equation}
\sum_{s=1}^{\infty}w_s\int_0^\infty
\frac{\theta^2 J_0^2(\theta)d\theta} {\left(\lambda_s^2-\theta^2\right)
\left(\lambda_m^2-\theta^2\right)}= \frac{3\pi}{2\lambda_m^2}.
\label{eq12}
\end{equation}
In order to get the solution for Eq.(\ref{eq12}) we will introduce an
integer odd function $f_1(z)$ the values of which in the points
$z=\lambda_s$ are equal
\begin{equation}
f_1(\lambda_s)=w_sJ_1(\lambda_s).  \label{eq13}
\end{equation}
Let us assume that at $\mid z\mid \rightarrow \infty$ $\; f_1(z)$
grows not faster than $\exp(z)$, then, in accordance with Cauchy theorem,
the function  $\left(f(z)/J_0(z)\right)$ can be expanded in a patial
fraction series
\begin{equation}
f_1(z)/J_0(z)=2z\sum_{n=1}^{\infty}w_n/\left(\lambda_n^2-z^2\right).
\label{eq14}
\end{equation}
Using (\ref{eq14}), and, also, multiplying Eq.(\ref{eq12}) by
$J_1(\lambda_m x)/J_1(\lambda_m)$, where $0<x<1$, and doing summation over
sub-index $m$, we will get
\begin{equation}
\int_0^\infty f_1(z)J_1(x z) dz=3\pi x/2, \; 0<x<1. \label{eq15}
\end{equation}
By multiplying (\ref{eq14}) by $zJ_1(xz)$ and integrating over $z$ from $0$
to  $\infty$, we will obtain (see the Appendix) at $x>1$:
\begin{equation}
L_1(x)=\int_0^\infty z f_1(z) J_1(xz) dz=0, \; x>1. \label{eq16}
\end{equation}

In this way, the set of linear algebraic equations  (\ref{eq12}) with
a complicated coefficients matrix that cannot be expressed via elementary
functions and can be calculated only numerically, has been reduced to two
integral equations (\ref{eq15}),(\ref{eq16}). Having determined the kind
of function $f_1(z)$, there is no need in calculating the sum (\ref{eq11}),
since
\begin{equation}
\Lambda=\sum_s w_s / \lambda_s^2=\lim_{z\to 0}{f_1(z)/\left(2zJ_0(z)\right)}.
\label{eq17}
\end{equation}

The method of solving the dual integral equations of the type
(\ref{eq15},\ref{eq16}) on the base of the Mellin transformation, as well as
the property of Cauchy-type integrals, can be found in \cite{r7}.
The brief summery of their solutions is given in \cite{r8}. We shall dwell
briefly  on a simpler method of resolving the system, because  it will be used
in Sec.3 for the analysis of the infinitely thick wall case.

Since $f_1(z)$ is the odd function it can be represented in the form
\begin{equation}
f_1(z)=\int_0^{\infty} \sin(zt)
\eta(t) dt. \label{eq18}
\end{equation}
Substituting this expression in Eq.(\ref{eq16}) we obtain such integral
equation for $\eta(t)$:
$$
\int_x^{\infty}\frac{\eta(t) dt}{\sqrt{t^2-x^2}}=0,\; x>1.
$$
The solution of this equation is $\eta(t)=0$ for $t>1$.
Consequently, any function of the type
\begin{equation}
f_1(z)=\int_0^1 \sin(zt)
\eta(t) dt \label{eq19}
\end{equation}
satisfies Eq.(\ref{eq16}).
Substituting (\ref{eq19}) into
(\ref{eq15}), we obtain the first kind Volterra equation Abelian type
\begin{equation}
\int_0^x
\frac{t\eta(t)dt}{\sqrt{x^2-t^2}}=\frac{3\pi}{2}x^2, \ 0<x<1, \label{eq20}
\end{equation}
the solution for which can be found in the analytical form. Omitting the
intermediate formulae, we shall give the final expression for the
function $f_1(z)$
\begin{equation}
f_1(z)=\frac{6}{z^2}\left\{\sin(z)-z \cos(z)\right\}
\approx 2z\left(1-\frac{z^2}{10}\right)
\label{eq21}
\end{equation}
The normalized coupling coefficients, as follows from (\ref{eq17}),
is equal to $\Lambda=1$. Since $w_s=f(\lambda_s)/J_1(\lambda_s)$, then, from
(\ref{eq9}), we will obtain
\begin{equation}
E_r^{(-)}(r)=\frac{E_{0,p,q}^{(1)}(r=0)-E_{0,p,q}^{(2)}(r=0)}{\pi}
\frac{r}{\sqrt{a^2-r^2}}
\label{eq22}
\end{equation}

Thus, on the base of a rigorous electrodynamic description of the two cavity
coupling system we are the first to prove, by the way of the limit
transition $a\rightarrow 0$, the correctness of the equations
formulated in the papers \cite{r4,r5,r6} on the basis of the quasi-static
approximation, and to obtain the expression for the tangential electric field
on the hole.

\subsection{The case of small, though finite, values of coupling hole radius}

The above method presents the opportunity to obtain analytical expressions for
the normalized coupling coefficients with an accuracy on the order of
$(a/\lambda)^2$.  If $a/\lambda$ is small, though finite, then, the
coefficients  $w_s$ in (\ref{eq10}) will be dependent on the hole radius
value $a$:  $w_s=w_s(a)$. Let's introduce the function of two variables:
\begin{equation}
\psi(a,z)=2zJ_0(z)\sum_{n=1}^{\infty} \frac{w_n(a)}{\lambda_n^2-z^2}
\label{eq23}
\end{equation}
We will assume that relative to the variable $z$ the function $\psi(a,z)$
will obey the conditions formulated in Subsec.2.1. Using the technique,
similar to that described in Subsec.2.1, the set (\ref{eq10}) can be
reduced to:
\begin{eqnarray}
\sum_{\ell=1}^{\infty}\frac{\theta_\ell
J_1\left(x\theta_\ell\right)}{\chi_\ell}  \psi\left(a,\theta_\ell\right)=0,
\; 1<x<b/a,
\label{eq24} \\
\pi \frac{a}{b}\sum_{\ell=1}^{\infty}\frac{J_1\left(x\theta_\ell\right)}
{\chi_\ell}
 \psi\left(a,\theta_\ell\right) R_\ell=
\frac{3\pi J_1\left(x\Omega_\ast\right)}{\Omega_\ast
J_0\left(\Omega_\ast\right)}, \; 0<x<1.
\label{eq25}
\end{eqnarray}

Letting $a\rightarrow 0$ in Eqs.(\ref{eq24}),(\ref{eq25}), we derive a set of
equations (\ref{eq15},\ref{eq16}), and, consequently, $\psi(0,z)=f_1(z)$,
where $f_1(z)$ is determined by Eq.(\ref{eq21}). Let's represent
$\psi(a,z)$ in the form
\begin{equation}
\psi(a,z)=\psi(0,z)+a^2 \varphi(a,z), \label{eq26}
\end{equation}
where $\varphi(a,z)$ is a function which has the same conditions imposed upon
that $\psi(a,z)$ does.

From (\ref{eq24}),(\ref{eq25}) it follows that $\varphi(0,z)$ satisfies
the following equations
\begin{eqnarray}
\int_{0}^{\infty} \theta
J_1\left(x\theta\right) \varphi\left(0,\theta\right)d\theta=0,
\ x>1,     \label{eq27} \\
\int_{0}^{\infty} J_1\left(x\theta\right) \varphi\left(0,\theta\right)
d\theta=F(x),\ 0<x<1, \label{eq28}
\end{eqnarray}
where
$$ F(x)=\lim_{a\to 0}
{\frac{1}{a^2}\left[ \frac{3\pi J_1\left(x\Omega_\ast\right)}{\Omega_\ast
J_0\left(\Omega_\ast\right)}- \pi
\frac{a}{b}\sum_{\ell=1}^{\infty}\frac{J_1\left(x\theta_\ell\right)}
{\chi_\ell} \psi\left(0,\theta_\ell\right) R_\ell \right]}.
$$
The coefficients $R_\ell$ can be represented as:
$$
R_\ell=1+\frac{\Omega^2}{2\theta_{\ell}^2}+\hat R_\ell.
$$
It can be shown that the following estimations are true
$$
\pi \frac{a}{b}\sum_{\ell=1}^{\infty}\frac{J_1\left(x\theta_\ell\right)}
{\chi_\ell}
\psi\left(0,\theta_\ell\right) =\frac{3\pi}{2}x+O\left(a^3\right),
$$
$$
\pi \frac{a}{b}\sum_{\ell=1}^{\infty}\frac{J_1\left(x\theta_\ell\right)}
{\chi_\ell}
\psi\left(0,\theta_\ell\right) \hat R_\ell =O\left(a^3\right).
$$
Then
$$
F(x)=\frac{3\pi}{8}x\left[\frac{\Omega_\ast^2-\Omega^2}{a^2}-
\frac{2\Omega_\ast^2-\Omega^2}{4a^2}x^2\right].
$$
The solution of Eqs.(\ref{eq27},\ref{eq28}) has the form
$$
\varphi(0,z)=\frac{\Omega_\ast^2-\Omega^2}{4a^2}f_1(z)-
\frac{2\Omega_\ast^2-\Omega^2}{2a^2}f_2(z),
$$
where $f_1(z)$ is determined by the formula (\ref{eq21}), while $f_2(z)$ is
\begin{equation}
f_2(z)=\frac{\left(3z^2-6\right)\sin(z)-z\left(z^2-6\right)\cos(z)}{z^4}
\approx\frac{z}{5}. \label{eq33}
\end{equation}
The function $\psi(a,z)\approx\psi(0,z)+a^2\varphi(0,z)$, accurate to the
order $(a/\lambda)^2$, has the form
\begin{equation}
\psi(a,z)\approx\left(1+\frac{\Omega_\ast^2-\Omega^2}{4}\right)f_1(z)-
\frac{2\Omega_\ast^2-\Omega^2}{2}f_2(z). \label{eq34}
\end{equation}
The normalized coupling coefficients  $\Lambda$ is determined by the
relationship
$$
\Lambda=J_0\left(\frac{a}{b}\lambda_q\right)
\psi\left(a,\frac{a}{b}\lambda_q\right)
/\left(2\frac{a}{b}\lambda_q\right) \approx
$$
\begin{equation}
\approx 1-\frac{1}{5}\left(\frac{a}{b}\lambda_q\right)^2-
\frac{3}{20}\left(\frac{a}{c}\omega_{q,p}\right)^2-
\frac{1}{20}\left(\frac{a}{c}\omega\right)^2. \label{eq35}
\end{equation}
For the case $\omega\approx\omega_{q,p}$  the expression (\ref{eq35}) agrees
with that for the generalized polarizability, obtained in \cite{r9}
at $b\rightarrow\infty$ via the variation technique. Note that the
expression (\ref{eq35}) is true for the frequency
$\omega$ that is not close to the resonant frequencies of the non-fundamental
modes of closed cavities:
$\omega\not=\omega_{m,n}, \mbox{ if } \ (m,n)\not=(q,p)$.

Knowing $\psi(a,z)$, and, consequently,  $w_s(a)=\psi\left(a,\lambda_s\right)
/J_1\left(\lambda_s\right)$, the form of the tangential electric field around
the hole can be reconstructed
$$
E_r^{(-)}(r)=\frac{E_{0,p,q}^{(1)}(r=0)-E_{0,p,q}^{(2)}(r=0)}{\pi}\times
$$
\begin{equation}
\times\left\{\left[1-\frac{1}{4}\left(\frac{a}{b}\lambda_q\right)^2+
\frac{\Omega_\ast^2-2\Omega^2}{12}\right] \frac{r}{\sqrt{a^2-r^2}}+
\frac{2\Omega_\ast^2-\Omega^2}{6}\frac{r}{a}\sqrt{1-
\left(\frac{r}{a}\right)^2} \right\}.
\label{eq36}
\end{equation}
Since in our approach the case $\Omega\rightarrow0$ corresponds to the
quasi-static method of field calculation, then, the formulae
(\ref{eq35},\ref{eq36}) at  $\Omega=0$ present the solution for the
appropriate quasi-static problem up to the second-order approximation in
$(a/b)$ in the case when the "far" field (see, \cite{r5},\cite{r6}) is not
homogeneous.

\section{Infinitely thick wall case}

The above analytical results pertain to the case of infinitely thin wall,
when the singularity at the rim of the hole has the form $(r-a)^{-1/2}$.
It is of interest to consider the possibility to apply the above method
to a non-zero thickness wall, when the singularity at the rim of the hole
has the form $(r-a)^{-1/3}$ \cite{r10}. The simplest problem in this class,
although representing an important application, is research into the coupling
of a cylindrical cavity with a co-axial cylindrical waveguide of the radius
$a<b$. Such a system can be studied upon the base the above equations in
the limit case  $t\rightarrow\infty$:
\begin{equation}
\epsilon_pZ_{q,p}a_{q,p}=-\omega_{p,q}^{2}\frac{4}{3\pi
J_1^2(\lambda_q)} \frac{a^3}{b^2d}
\Lambda a_{q,p}, \label{eq37}
\end{equation}
where the normalized coupling coefficients at $a\rightarrow0$ is equal to
\begin{equation}
\Lambda=\sum_{s=1}^{\infty}w_s/\lambda_s^2, \label{eq38}
\end{equation}
while $w_s$ are the equation solutions
\begin{equation}
\frac{w_m}{2\lambda_m}+\sum_{s=1}^{\infty}w_s\int_0^\infty
\frac{\theta^2 J_0^2(\theta)d\theta} {\left(\lambda_s^2-\theta^2\right)
\left(\lambda_m^2-\theta^2\right)}= \frac{3\pi}{\lambda_m^2}.
\label{eq39}
\end{equation}
The set (\ref{eq39}) is different from the above-studied (\ref{eq12})
in additional addends in the diagonal matrix elements.
Introducing a function of the type (\ref{eq14}), we obtain the following
set of equations:
\begin{equation}
\int_0^\infty z f(z) J_1(xz)dz=0, \ x>1. \label{eq40}
\end{equation}
\begin{equation}
\sum_{m=0}^{\infty}\frac{f\left(\lambda_m\right)J_1\left(x\lambda_m\right)}
{\lambda_mJ_1^2\left(\lambda_m\right)}+
\frac{1}{2}\int_0^\infty f(z)J_1(xz)dz=\frac{3\pi x}{2}, \ 0<x<1.
\label{eq41}
\end{equation}
As indicated above, from (\ref{eq40}) it follows that $f(z)$ should be sought
in the form of (\ref{eq19}), then, from (\ref{eq41}) we get the fact that
$\eta(t)$ has to be the solution of the Fredholm equation of the second
kind
\begin{equation}
\eta(u)+\frac{4}{\pi}\int_0^1 \eta(t)\sum_{m=1}^{\infty}
\frac{\sin(\lambda_mt)\sin(\lambda_mu)}{\lambda_m
J_1^2\left(\lambda_m\right)}dt= 12u. \label{eq42} \end{equation}
Since the kernel of this integral equation is degenerate, then its solution
have the appearance
\begin{equation}
\eta(u)=12u-\frac{4}{\pi}\sum_{m=1}^{\infty} \frac{\sin(\lambda_m u)}
{\lambda_m J_1^2\left(\lambda_m\right)} c_m, \label{eq43}
\end{equation}
where
$$
c_m=\int_0^1 \eta(t)\sin(\lambda_m t) dt
$$
the constant coefficients which are the solution of the infinite linear set
of equations to be easily obtained by way of the appropriate integration
(\ref{eq43}).  Since $c_m=f\left(\lambda_m\right)=w_m
J_1\left(\lambda_n\right)$ (see (\ref{eq19})), this system can be represented
as:
$$
w_m+\frac{2}{\pi}\sum_{n=1}^{\infty}
\frac{w_n}{\lambda_n J_1\left(\lambda_n\right)J_1\left(\lambda_m\right)}
\left[\frac{\sin(\lambda_n-\lambda_m)}{\lambda_n-\lambda_m}-
\frac{\sin(\lambda_n+\lambda_m)}{\lambda_n+\lambda_m}\right]=
$$
\begin{equation}
=\frac{12\left\{\sin(\lambda_m)-\lambda_m\cos(\lambda_m)\right\}}
{\lambda_m^2J_1\left(\lambda_m\right)}. \label{eq44}
\end{equation}

Thus, for fields with the singularity at the rim of the hole of the type
$(r-a)^{-1/3}$ we have obtained an analogous system (\ref{eq44}) instead
of the initial linear set of equations (\ref{eq39}). Comparing (\ref{eq44})
and (\ref{eq39}), we can see that in the system (\ref{eq39}) the matrix
coefficients are expressed through the improper integrals, whereas in the
modified set  (\ref{eq44}) they are determined by well studied functions.
This considerably facilitates both analytical studies and numerical
simulation.

We have carried out numerical calculations of the normalized coupling
coefficient $\Lambda$ determined by the formula (\ref{eq38}) both on the base
of (\ref{eq39}) and (\ref{eq44}). For a 200*200 matrix, the calculations based
on (\ref{eq39}) gives $\Lambda=0.85835$, based on (\ref{eq44}) ---
$\Lambda=0.85854$, that is the results agree up to $2\times10^{-4}$.
This result confirms the correctness of the above analytical method.
Note that the obtained $\Lambda$-value corresponds to the results of a purely
static analysis  \cite{r11} carried out on the base of the variation
technique for the infinite plane with a cylindrical hole.

\section{Conclusion}

On the base of our method of reduction of the infinite linear algebraic
equation set to dual integral equations, we obtained, in different limited
cases, the rigorous analytical solutions regarding the two-cavity coupling
problem. Alongside with general theory significance, the obtained solutions
are of applied interest, since they can be used for a better convergence
of the original equation solution (\ref{eq4},\ref{eq5}), which are true
for arbitrary dimensions of the coupling hole.

\section{Acknowledgment}

The author wishes to thank Academician V.A.Marchenko and Professor
V.I.Kurilko for discussion relating  results of this work.

\newpage
\section{Appendix}
\def\theequation{A.\arabic{equation}}
\setcounter{equation}0

If $f_1(z)$ can be representable as series (\ref{eq13}), then
\begin{equation}
L_1(x)=\int_0^\infty z J_1(xz)f_1(z) dz=2\sum_{n=0}^{\infty}w_n G_n(x).
\label{q2}
\end{equation}
where
$$ G_n(x)=\int_0^\infty \frac{z^2
J_1(xz)J_0(z)dz}{\lambda_n^2-z^2}= $$
\begin{equation}
=-\int_0^\infty
J_1(xz) J_0(z)dz + \lambda_n^2 \int_0^\infty \frac{
J_1(xz)J_0(z)dz}{\lambda_n^2-z^2}= -G^{(1)}(x)+\lambda_n^2 G_n^{(2)}(x).
\label{q3}
\end{equation}
It is easy to demonstrate (see \cite{r8}), that the integral $G^{(1)}$
at $x>1$ is equal $G^{(1)}=1/x$.
Let's consider the integral  $G_n^{(2)}(x)$ at $x>1$:
\begin{equation}
G_n^{(2)}(x)=\int_0^\infty \frac{ J_1(xz)J_0(z)dz}{\lambda_n^2-z^2}.
\end{equation}
Having made the substitution $y=xz$, we obtain
\begin{equation}
G_n^{(2)}(x)=\alpha\int_0^\infty \frac{ J_1(y) J_0(\alpha
y) dy}{\lambda_n^2-\alpha^2 y^2}, \label{q5}
\end{equation}
where $\alpha=1/x,\  0<\alpha<1$.
Using the expansion
$$
J_1(y)=2yJ_0(y)\sum_{m=1}^\infty \frac{1}{\lambda_m^2-y^2}.
$$
(\ref{q5}) will take the form
$$
G_n^{(2)}(x)=2\alpha\sum_{m=1}^\infty \int_0^\infty \frac{y J_0(y) J_0(\alpha
y) dy}{\left(\lambda_n^2-\alpha^2 y^2\right)\left(\lambda_m^2-y^2\right)}=
$$
$$
=2\alpha\sum_{m=1}^\infty \frac{1}{\alpha^2\lambda_m^2-\lambda_n^2}
\left[ \alpha^2
\int_0^\infty \frac{ y J_0(y) J_0(\alpha y) dy}{\left(\lambda_n^2-\alpha^2
y^2\right)} -
\int_0^\infty \frac{ y J_0(y) J_0(\alpha y) dy}{\left(\lambda_m^2-y^2\right)}
\right]=
$$
\begin{equation}
=2\alpha\sum_{m=1}^\infty \frac{1}{\alpha^2\lambda_m^2-\lambda_n^2}
\left[\alpha^2 P_n^{(1)}-P_m^{(2)}\right]. \label{q6}
\end{equation}
Calculation of the integrals under consideration is based on such
expansion (see\cite{r8}),
$$
\frac{\pi}{4} \frac{J_0(\alpha\beta z)}{J_0(z)}
\left[J_0(z)Y_0(\beta z)-J_0(\beta z)Y_0(z)\right]=
$$
\begin{equation}
=\sum_{s=1}^\infty \frac{J_0\left(\alpha \beta \lambda_s\right)
J_0\left(\beta \lambda_s\right)}{J_1^2\left(\lambda_s\right)
\left(z^2-\lambda_s^2\right)}, \label{q8}
\end{equation}
where $0<\alpha<1, \ 0<\beta<1$.

Let's consider (\ref{q8}) at $z=\lambda_m/\beta$
$$
\sum_{s=1}^\infty \frac{J_0\left(\alpha \beta \lambda_s\right)
J_0\left(\beta \lambda_s\right) \beta^2}{J_1^2\left(\lambda_s\right)
\left(\lambda_m^2-\beta^2\lambda_s^2\right)}=\frac{1}{2}
\frac{J_0\left(\alpha  \lambda_m\right)}{J_1\left(\lambda_m\right)\lambda_m}
$$
or
$$
\pi\beta \sum_{s=1}^\infty \frac{J_0\left(\alpha \beta \lambda_s\right)
J_0\left(\beta \lambda_s\right) \beta
\lambda_s}{\left\{\pi\lambda_s J_1^2\left(\lambda_s\right)/2\right\}
\left(\lambda_m^2-\beta^2\lambda_s^2\right)}=
\frac{J_0\left(\alpha  \lambda_m\right)}{J_1\left(\lambda_m\right)\lambda_m}
$$
At $\beta\rightarrow0$, we have
\begin{equation}
P_m^{(2)}=
\int_0^\infty \frac{ y J_0(y) J_0(\alpha y) dy}{\left(\lambda_m^2-y^2\right)}=
\frac{J_0\left(\alpha  \lambda_m\right)}{J_1\left(\lambda_m\right)\lambda_m}.
\label{q9}
\end{equation}

The value  $P_n^{(1)}=0 $  can be obtained from (\ref{q9}) at
$\lambda_m\rightarrow\lambda_n/\alpha$.  Then
$$
G_n^{(2)}(x)=2\alpha\sum_{m=1}^\infty  \frac{ J_0\left(\alpha
\lambda_m\right)}{\left(\lambda_n^2-\alpha^2\lambda_m^2\right)
\lambda_m J_1\left(\lambda_m\right)}.
$$
From the expansion
$$
\frac{J_0(\alpha z)}{J_0(z)}=1-2 z^2\sum_{m=1}^\infty
\frac{J_0\left(\alpha \lambda_m\right)}{\lambda_m J_0\left(
\lambda_m\right)}\frac{1}{z^2-\lambda_m^2}.
$$
at $z=\lambda_n/\alpha$ we have
$$
\sum_{m=1}^\infty \frac{J_0\left(\alpha \lambda_m\right)}
{\lambda_m
J_0\left(\lambda_m\right)}\frac{1}{\lambda_n^2-\alpha^2\lambda_m^2} =
\frac{1}{2\lambda_n^2}.
$$
Consequently,
$$
G_n^{(2)}(x)=\frac{\alpha}{\lambda_n^2}=\frac{1}{x\lambda_n^2}.
$$
and
$$
G_n(x)=\int_0^\infty \frac{z^2 J_1(xz)J_0(z)dz}{\lambda_n^2-z^2}=
-G^{(1)}(x)+\lambda_n^2 G_n^{(2)}(x)=-\frac{1}{x}+\frac{1}{x}=0.
$$
Finally,
\begin{equation}
L_1(x)=\int_0^\infty z J_1(xz)f_1(z) dz=2\sum_{n=0}^{\infty}w_n G_n(x) = 0,
\ x>1,
\label{q11}
\end{equation}
which is what needed to be proven.

\end{document}